%% file: fse.tex
\documentclass[12pt]{article}
\usepackage{amsmath,epsfig,amssymb,eufrak}

\date{}

\newcommand{\e}{{\rm{e}}}

\input macros.sty      
\input eqalign.sty     

\begin{document}

\begin{titlepage}

\title{
  {\vspace{-0cm} \normalsize
  \hfill \parbox{40mm}{DESY 04-041\\
                       SFB/CPP-04-08\\March 2004}}\\[30mm]
Finite size effects of a pion matrix element}
  \author{M.\ Guagnelli$^{a}$, K.\ Jansen$^{b}$, F.\ Palombi$^{a,c}$,\\ 
   R.\ Petronzio$^{a}$, A.\ Shindler$^{b}$ and I.\ Wetzorke$^{b}$\\
\\
   {Zeuthen-Rome (ZeRo) Collaboration}\\
\\ 
  {\small $^a$ Dipartimento di Fisica, Universit\`a di Roma 
        {\em Tor Vergata}}\\ 
  {\small and INFN, Sezione di Roma II,} \\
  {\small Via della Ricerca Scientifica 1, I-00133 Rome, Italy} \\
  {\small $^b$ NIC/DESY Zeuthen, Platanenallee 6, D-15738 Zeuthen,
   Germany}\\
  {\small $^c$ {\em E.~Fermi} Research Center,
   c/o Compendio Viminale, pal.~F, I-00184 Rome, Italy} \\
}

\maketitle

\begin{abstract}
We investigate finite size effects of the 
pion matrix element of the non-singlet, 
twist-2 operator corresponding to the average
momentum of non-singlet quark densities.
Using the quenched approximation, 
they come out to be surprisingly large when compared to the 
finite size effects of the 
pion mass.
As a consequence, 
simulations of corresponding nucleon matrix elements could 
be affected by finite size effects even stronger which 
could lead to serious 
systematic uncertainties in their evaluation.
\vspace{0.8 cm}
\noindent
\end{abstract}

\end{titlepage}

\section{Introduction}

A complete description of moments of parton distribution functions 
(PDF) from first
principles is still missing. In order to provide reliable 
numerical determinations, it is necessary to control all  
systematic uncertainties which include 
the continuum limit, non-perturbative renormalization, chiral extrapolation,
finite size effects and quenching.
Several studies on hadron matrix elements of twist-2 operators have been
performed since now \cite{{ref:proton_LHPC},{ref:proton_QCDSF}}.
In \cite{ref:proton_QCDSF} a study of the continuum limit employing 
a perturbative renormalization shows small cutoff effects. 
Applying a non-perturbative renormalization \cite{ref:x_running},
slightly reduces an observed discrepancy between theory and 
experiment \cite{ref:nonpert_proc}.
First computations in the unquenched case \cite{ref:proton_LHPC}, 
at a rather high value of the quark mass,
seems to indicate no quantitative differences with the quenched case.
In \cite{ref:chiral_ext} it has been suggested that the discrepancy 
between theory and experiment
could be explained by a rather strong dependence of the matrix elements
on the quark mass in the chiral region when a very small quark
mass is entered.
The simulations performed since now are carried out at values of the quark
mass too far away from the physical point to test this 
anticipated mass dependence.
One problem that to our knowledge has not been faced in a systematic manner
is how the finite size of the lattice
can influence such computations.

In this letter we present a study of the finite size effects (FSE) of the 
quenched pion matrix element of the twist-2 operator corresponding 
to the average momentum of non-singlet quark densities.
Preliminary results were presented at the last lattice conference
\cite{ref:fse_proc}.

It has already been noticed some time 
ago \cite{{ref:fse_dyn}} and more recently in 
\cite{{ref:fse_wup},{ref:fse_qcdsf}}, 
that
FSE expecially in the unquenched case can be a serious source of 
systematic error.
In this letter we want to emphasize that even in the quenched case, FSE for
pion 3-point functions involving twist-2 operators, can be much 
larger than the FSE 
for the hadron masses.
Thus, there is a clear warning that for the unquenched case and moreover
for nucleon matrix elements such FSE should be studied very carefully 
in order to avoid an underestimation of this systematic effect. 

\section{Setup and basic definitions}

The moments of parton distribution functions (PDF) $\langle x^{(N-1)} \rangle $
are related to matrix elements of leading twist $\tau$ ($\tau=$dim-spin) 
operators ${\cal O}$ of given spin, between hadron states $h(p)$ 
\bea \hspace*{-5mm} 
\langle h(p)|{\cal O}_{\mu_1 \ldots \mu_N}|h(p) \rangle 
&=& M^{(N-1)}(\mu) p_{\mu_1} \cdots p_{\mu_n} \nn \\ 
&& + {\rm{terms~}} \delta_{\mu_i \mu_j} 
\eea 
\be \hspace*{-7mm} \langle x^{(N-1)} \rangle (\mu) = M^{(N-1)}(\mu = Q)\; ,
\ee
where the renormalization scale $\mu$ is identified with the momentum transfer
$Q$ in deep inelastic scattering (DIS).

Our setup of lattice QCD is on a hyper-cubic 
euclidean lattice with 
spacing $a$ and size
$L^3 \times T$.
We impose periodic boundary conditions in the spatial directions and Dirichlet
boundary conditions in time, as they are used to formulate the Schr\"odinger
functional (SF) \cite{ref:sf_fond} 
(we refer to these references for unexplained notations).
In this paper we will consider homogeneous boundary conditions, 
where the spatial components
of the gauge potentials at the boundaries and also the fermion boundary source
fields are set to zero. In this case the Schr\"odinger functional partition 
function can be written as
\cite{ref:me_sf} 

\be
\cZ = \langle i_0 | \e^{-T\Ham} \Pgauge|i_0 \rangle
\ee
where the initial and final states $|i_0 \rangle$ carry the quantum 
numbers of the vacuum.
We have studied the matrix elements between charged pion states of the 
following operator
\be
\cO_{44}(x) = \frac{1}{2} \bar\psi(x) \Big[ \gamma_4 \lrD_4 - \frac{1}{3}
\sum_{k=1}^3 \gamma_k \lrD_k \Big] \psi(x)\; .
\label{O44}
\ee
We indicate with $\zeta$ and $\bar\zeta$ 
(and the corresponding $\zeta'$ and $\bar\zeta'$)
a flavour doublet. As discussed in \cite{ref:Oa_imp} we recall that
$\zeta$, $\bar\zeta$, $\zeta'$ and $\bar\zeta'$ are functional derivatives
with respect to the boundary source fields.

The states with charged pion quantum numbers 
in the Schr\"odinger functional are the dimensionless fields
\be
\Source = \frac{a^6}{L^3} \sum_{\by,\bz} \bar\zeta(\by)\gamma_5 \tau^+ 
\zeta(\bz) \qquad
\Source' = \frac{a^6}{L^3} \sum_{\bu,\bv} 
\bar\zeta'(\bu)\gamma_5 \tau^- \zeta'(\bv)
\label{eq:source_q}
\ee
where $\tau^{\pm}=\frac{1}{\sqrt{2}}(\tau^1 \pm i \tau^2)$ and 
$\tau^k$ with $k=1,2,3$ 
are the usual Pauli matrices.
The correlation function to extract the matrix element is
\be
f_{44}(x_0) = \langle \Source \cO_{44}(x) \Source' \rangle\; . 
\ee
The Wick contractions of this correlation function contain also 
a disconnected piece that we neglect consistently with the fact 
that we will do quenched simulations.
For normalization purposes it is important to define
\be
f_1 = -\frac{1}{2} \langle \Source \Source' \rangle\; .
\ee
The correlation functions $f_{44}$ and $f_1$ have the following 
quantum mechanical
representations
\be
f_{44}(x_0) = 
\cZ^{-1} \langle i_{\pi} | \e^{-(T-x_0)\Ham} \Pgauge \cO_{44}(x) \e^{-x_0 \Ham} \Pgauge |i_{\pi} \rangle
\ee
\be
f_1 = \cZ^{-1}\frac{1}{2}\langle i_{\pi} |\e^{-T \Ham} \Pgauge |i_{\pi} \rangle
\; . 
\ee
Inserting a complete set of eigenstates of the Hamiltonian and retaining
only the first non-leading corrections we have
\be
f_{44}(x_0) \simeq \rho^2 
\langle 0,\pi| \cO_{44}(x) | 0,\pi \rangle \e^{-m_{\pi}T} 
\{1 + \eta_{\cO_{44}}^\pi \e^{-x_0\Delta} + 
\eta_{\cO_{44}}^\pi \e^{-(T-x_0)\Delta}\}
\ee
\be
f_1 \simeq \rho^2 \e^{-m_{\pi}T}\; .
\ee
The matrix element we are interested in, neglecting contributions 
from excited states, 
can then be extracted from the plateau value of the following ratio:
\be
\frac{f_{44}(x_0)}{f_1} = \langle 0,\pi| \cO_{44}(x) | 0,\pi \rangle \; .
\ee
In order to relate this numerically computed ratio with the corresponding
continuum operators in the Minkowski space, we need a suitable normalization
factor
\be
\langle x \rangle = \frac{2 \kappa}{m_\pi}
\langle 0,\pi| \cO_{44}(x) | 0,\pi\rangle
\; ,
\label{norm}
\ee
where we always take for $m_\pi$ the infinite volume pion mass
$m_\pi(L=\infty)$, summarized in table \ref{table:fit_par}, in order to
disentangle the FSE of the pion mass and $\langle x \rangle$ itself.
Furthermore, we consider only the bare matrix elements without renormalization
factor, since the FSE remain unchanged when employing the appropriate
$Z$ factors.

Using the procedure of ref.~\cite{ref:me_sf} we have extracted 
the plateau values for 
the effective pion mass and the 3-point function with a 
systematic relative error
coming from the excited states of 0.1\%, respectively 0.4\%, well below 
the statistical
accuracy of our computations. 
Technical details can be found in a forthcoming publication 
\cite{ref:x_pion}.

\section{Numerical details and results}


\begin{table}[t]
\centering
\begin{tabular}{|c|c|c|c|}
\hline
$L/a$ & $T/a$ & $N$ \\\hline\hline
12 & 36 & 3200 \\\hline
14 & 36 & 1600 \\\hline
16 & 36 & 800  \\
16 & 42 & 800  \\\hline
24 & 36 & 800  \\
24 & 42 & 800  \\\hline
32 & 36 & 400  \\
32 & 56 & 185  \\\hline
\end{tabular}
\caption{\footnotesize 
Parameters of our simulation points. $N$ denotes the number of measurements
taken into account.}
\label{table:parameters}
\end{table}

We have performed a set of quenched simulations at fixed 
lattice spacing $a=0.079$ fm ($\beta=6.1$),
for three values of the quark mass 
($\kappa$ = 0.1340, 0.1345, 0.1350), and several
lattice sizes, using the non-perturbatively improved 
clover action \cite{ref:c_sw}.
For some simulations we have also used two different time 
extents of our lattice
($T$, $T'$, with $T' < T$).
Our simulation parameters are summarized in table \ref{table:parameters}.
Using $r_0=0.5$ fm to fix the scale \cite{ref:r0}, 
we have lattice sizes varying between
$0.9$ fm and $2.5$ fm, and values of $z = m_\pi L$ varying between 
$2.5$ and $11.2$.


\begin{figure}
\begin{center}
\psfig{file=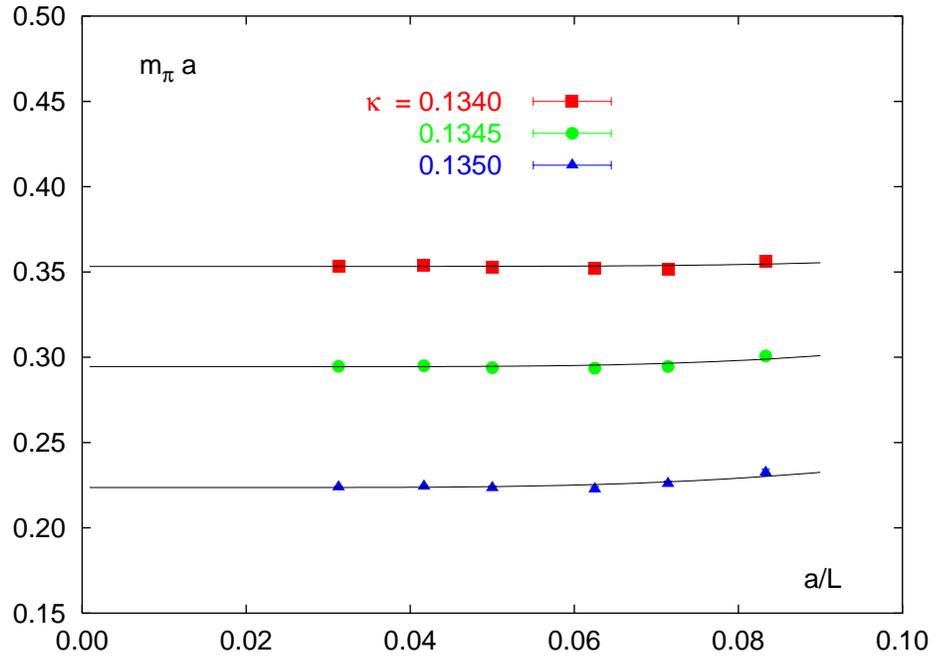, width=13cm}
\end{center}
\caption{ \label{fig:meff1}
Finite size dependence of the pion mass for three values of the quark masses.
The curves show the exponential fit of eq.~(\ref{exp_fit}) with $a_2=m_\pi$
fixed (see text).}
\end{figure}


\begin{figure}
\vspace{-0.5cm}
\begin{center}
\psfig{file=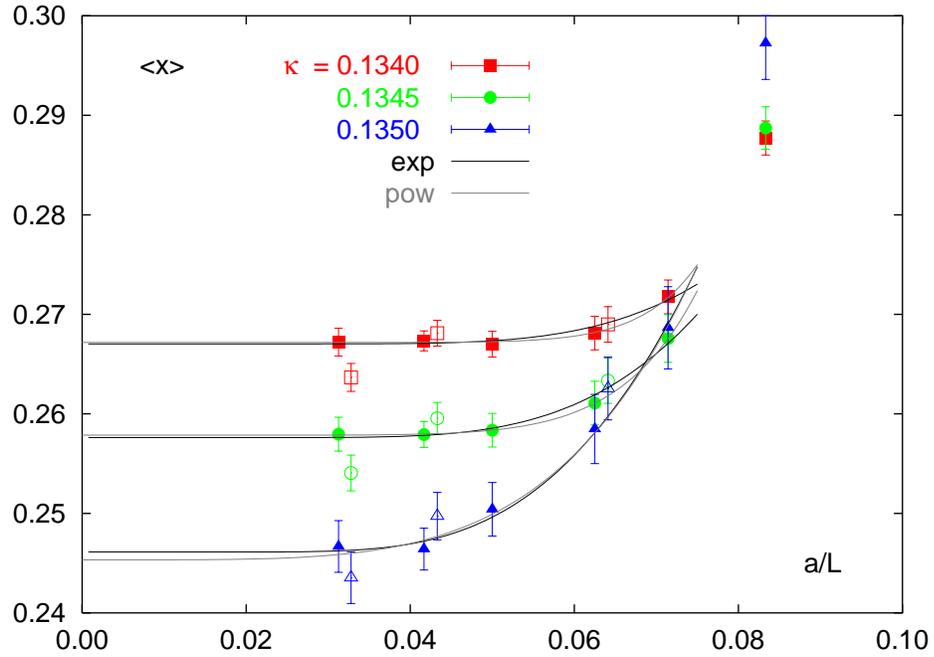, width=13cm}
\end{center}
\caption{ \label{fig:me1}
Finite size dependence of the twist-2 operator matrix element 
for three values of the quark masses.
The curves show both the exponential ($a_2=m_\pi$ fixed) and the power law
fits (see text). The open symbols
denote simulation results with lattices of the same spatial extent, 
but with a shorter time extent (see text). 
They are slightly displaced in $a/L$ for better visibility.}
\end{figure}


\begin{table}[t]
\centering
\begin{tabular} {|r||c|c||c|c|}
\hline
~ & $m_\pi(L=\infty)$ & $\chi^2/dof$ &
$\langle x \rangle (L=\infty)$ & $\chi^2/dof$ \\\hline
exp: $\kappa=0.1340$  & 0.3533(3) & 1.7 & 0.2672(7)  & 0.03\\
            $0.1345$  & 0.2946(3) & 0.6 & 0.2579(10) & 0.002\\
            $0.1350$  & 0.2239(4) & 0.6 & 0.2458(24) & 0.12\\\hline
exp$^*$: $\kappa=0.1340$  & 0.3533(3) & 2.2 & 0.2670(7)  & 0.12\\
            $0.1345$  & 0.2945(3) & 2.0 & 0.2576(9)  & 0.08\\
            $0.1350$  & 0.2237(4) & 1.6 & 0.2461(15) & 0.09\\\hline
pow: $\kappa=0.1340$  & 0.3533(3) & 1.6 & 0.2671(6)  & 0.03\\
            $0.1345$  & 0.2946(3) & 0.6 & 0.2579(11) & 0.002\\
            $0.1350$  & 0.2239(4) & 0.6 & 0.2453(29) & 0.17\\\hline
\end{tabular}
\caption{\footnotesize 
Results for the infinite volume limit of $m_\pi$ and $\langle x \rangle$
using the exponential fit (exp) of eq.~(\ref{exp_fit}), fixing $a_2=m_\pi$
(exp$^*$) and employing the power law fit of eq.~(\ref{pow_fit}) to describe
the FSE.}
\label{table:fit_par}
\end{table}

We have made fits of the data both with an exponential and power 
dependence on the lattice size $L$ (in the following equation we generically
indicate with $F(L)$ both $m_\pi$ and $\langle x \rangle$):
\be
{\rm exp:} \qquad F(L) = a_0 + \frac{a_1}{L^{3/2}}\exp\{-a_2L\}
\label{exp_fit}
\ee
\be
{\rm pow:} \qquad F(L) = a_0 + a_1 / L^{a_2}
\label{pow_fit}
\ee
The values of $m_\pi(L=\infty)$ as obtained from the different fits
are summarized in table \ref{table:fit_par} complemented by $\chi^2/dof$.
In fig.~\ref{fig:meff1} we show the size dependence of
$m_\pi$ for the three quark
masses. The solid curves show the exponential fits keeping $a_2=m_\pi$ fixed.
Also a power law fit describes the data rather well and leads to a comparable
number of the inifinite volume pion mass. 
For the pion mass in the unquenched case it is inferred that for very small 
lattices the FSE are described by a power law 
behaviour \cite{ref:pow_fse}. In this case
the source of the FSE is the squeezing of the hadron in a small volume.
Before the asymptotic (infinite volume) region is reached, 
there is a pre-asymptotic region where the
unquenched FSE are described by an exponential 
law \cite{ref:exp_fse} indicating that the
source of the FSE is the wrapping around the world of the virtual pions.
Our data obtained in the quenched approximation can not differentiate 
between these two cases. 

In fig.~\ref{fig:me1} we show the finite 
size dependence of $\langle x \rangle$ for
the three quark masses. As expected, the FSE turn out to be stronger for
lighter quark masses when keeping the box size fixed.
At present there are no predictions of the form of the 
FSE for a matrix element such as $\langle x \rangle$. 
As an ansatz, 
we have used the same fit functions 
as for the pion mass.
Again, the comparison of the matrix element with both 
fit functions does not allow to prefer either of the two fit
functions. Thus, carrying over the previous arguments for the pion
mass to the 3-point function, it is difficult to disentangle  
which is the
physical origin of the FSE of the matrix element.


\begin{figure}
\vspace{-1.0cm}
\begin{center}
\psfig{file=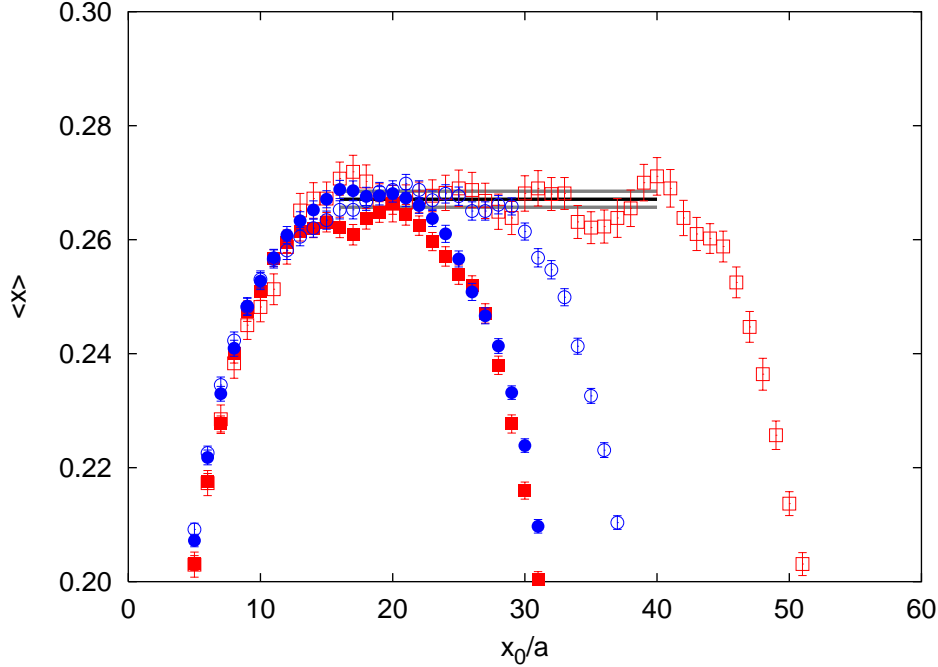, width=13cm}
\end{center}
\caption{ \label{fig:t_ext}
Time dependence of the 3-point function for $\kappa=0.1340$ coming
from the following lattice sizes:
$32^3 \times 56$, $32^3 \times 36$ (squares), $24^3 \times 42$ and
$24^3 \times 36$ (circles).}
\end{figure}

At this point we want to make a short remark about the importance of 
the physical extent of the plateau in time direction.
In fig.~\ref{fig:me1} the empty symbols show the values of 
$\langle x \rangle$ obtained with lattices of the same spatial extent, 
but with
a shorter time extent ($T' \sim 2.8 \rm{~fm} < T \sim 3.3-4.4 \rm{~fm}$).
Comparing $\langle x \rangle$ on the lattices $L/a=24$ and $L/a=32$ with
smaller time extent, it looks like there would still be
some FSE. This is, however, not the case and the explanation of this 
effect has a different source.
Since it takes around $1.3$ fm before all the contributions of the excited
states go below the accuracy of our computation, a time extent below or 
around $3$ fm
can be a source of an in-accurate determination of the plateau value, 
given the
small number of points available.
In fig.~\ref{fig:t_ext} we plot the time dependence of 
the 3-point function for $\kappa=0.1340$, coming
from lattices with different spatial extents ($L/a=24, 32$) and 
different time
extents ($T/a=36, 42, 56$). The two spatial extents should be large enough, 
for all our quark masses, to be free from FSE. 
Since the contributions of the excited states have already died out
at around $1.3$ fm, even if
the plateau is short it should give results consistent with the result
obtained from a longer time extent. From fig.~\ref{fig:t_ext} it is clear that
the disagreement between the two time extents is given by a statistical 
fluctuation that plays the role of a systematic error due to the short time
extent. In the longer time extent case too, there is a wiggling in the 
plateau
due to statistical fluctuations, but in the short time extent case the
plateau is not long enough to average away these statistical fluctuations.
This is also consistent with the fact that for the two different 
spatial sizes this fluctuation goes in opposite directions.
A warning we want to give is that if the number of points in the plateau is
only around $3-5$ there is the risk to have results biased by $2-3\;\sigma$,
that obviously are consistent with being a statistical fluctuation.


\begin{figure}
\vspace{-1cm}
\begin{center}
\psfig{file=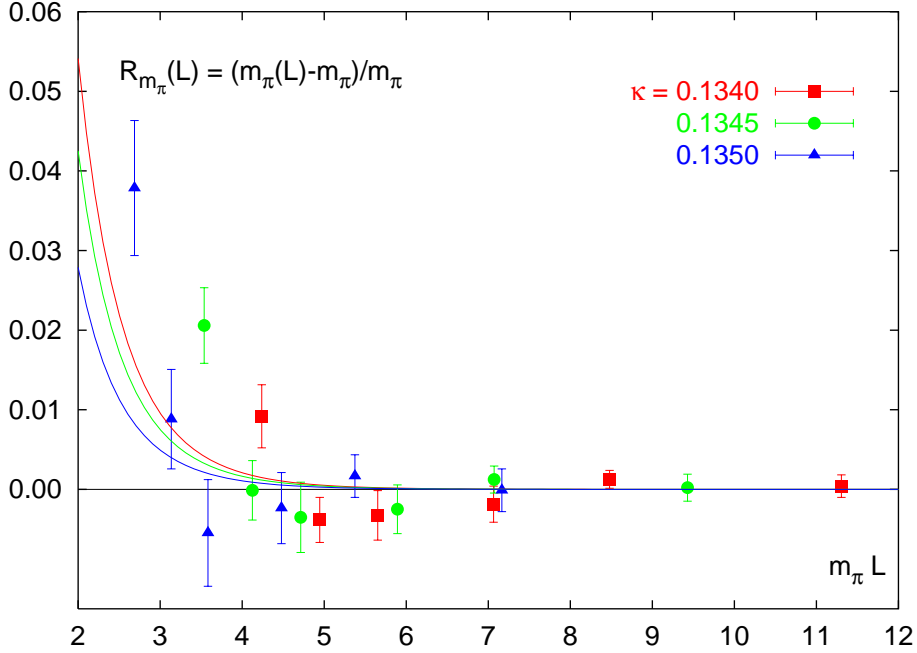, width=13cm}
\end{center}
\caption{ \label{fig:meff2}
Finite size dependence of $m_\pi$ as a function of $z=m_\pi L$.
The solid curves show the FSE for the pion mass described by
formula (\ref{ChPT}).}
\end{figure}


\begin{figure}
\vspace{-0.5cm}
\begin{center}
\psfig{file=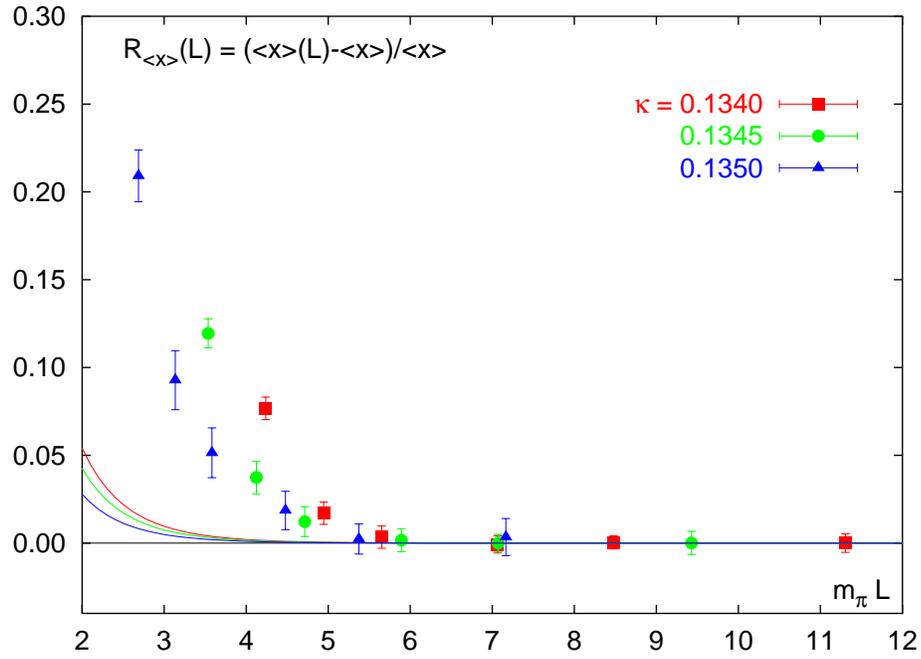, width=13cm}
\end{center}
\caption{ \label{fig:me2}
Finite size dependence of the twist-2 operator matrix element as a 
function of $z=m_\pi L$. The solid curves indicate the finite
size effects of the pion mass described by formula (\ref{ChPT})
for comparison.}
\end{figure}

It is however quite clear that the FSE for the 3-point functions are much
larger than for the pion mass. In order to make this statement 
more quantitative
we have plotted the relative systematic deviation

\be
R_F(L) = \frac{F(L) - F(L=\infty)}{F(L=\infty)} \qquad F(L) =
m_\pi(L),\langle x \rangle (L)
\ee
as functions of $m_\pi L$, where $m_\pi$ is the infinite volume
pion mass. Since the FSE for the pion mass are very small, it makes no
difference to study the dependence of the data as functions of 
$m_\pi(L) L$ or $m_\pi(L=\infty) L$. In fig.~\ref{fig:meff2} and \ref{fig:me2}
we see that both the data sets scale in a good way as function of
$m_\pi L$. Apart from the $L/a=12$ points the
data sets are in agreement with the formula given in
ref.~\cite{ref:gasser,ref:cola_durr}
to leading order of chiral perturbation theory
\bea
R_{m_\pi}(L)
&\simeq& \frac{3}{8 \pi^2} \left(\frac{m_\pi}{F_\pi}\right)^{\!2}
\left[
\frac{K_1(m_\pi L)}{m_\pi L} +
\frac{2 \; K_1(\sqrt{2} \; m_\pi L)}{\sqrt{2} \; m_\pi L}
\right]\nn\\
&\simeq& \frac{3}{4 (2 \pi)^{3/2}} \left(\frac{m_\pi}{F_\pi}\right)^{\!2}
\left[
\frac{e^{-m_\pi L}}{(m_\pi L)^{3/2}} +
\frac{2 \; e^{-\sqrt{2} \; m_\pi L}}{(\sqrt{2} \; m_\pi L)^{3/2}}
\right]\;,
\label{ChPT}
\eea
where we have interpolated the results for $m_\pi/F_\pi$ of
ref.~\cite{ref:garden} to our $\kappa$ values.

A possible explanation for this behaviour is the fact that the validity of the
p-expansion of chiral perturbation theory, used to determine formula
(\ref{ChPT}), is limited by the constraint $F_\pi L \gg 1$. For the lattice
size $L/a=12$ we find instead $F_\pi L \simeq 1$.

The interesting (and worrisome) result is that while the FSE for the pion mass
vanish for 
$m_\pi L \ge 4$, the FSE for $\langle x \rangle$ vanish only for   
$m_\pi L \ge 5.5$. At $m_\pi L \sim 4$ where the FSE for the pion
mass are consistent with zero, the 3-point function, due to the FSE, 
is overestimated by around $5\%$.

\section{Conclusions}

As suggested in \cite{ref:fse_dyn}, the breaking of the gauge center 
symmetry $Z_3$ can be responsible for larger FSE of the hadron masses
in the unquenched case compared to the quenched case. This effect, that
would lead to power law FSE, is dominant for intermediate volumes.
It is also well known that FSE are larger for the nucleon mass than for the
pion mass \cite{{ref:fse_dyn},{ref:fse_wup}}, and moreover, the lattice size
where the power law behavior disappears in favor of an exponential one is
larger for the nucleon than for the pion. There is then a danger
that the systematic uncertainty coming from FSE in the current unquenched
determinations of the nucleon matrix elements is severely underestimated.
At the moment there are no studies of FSE for 3-point functions
involving twist-2 operators between nucleon states, and to our knowledge, this
is the first study of FSE of matrix elements involving twist-2 operators.

We think, it would be important and urgent to have 
an analytical description of the FSE and in particular of
the size dependence of the twist-2 matrix elements between hadron states, 
along the line of the computations performed in the framework of 
chiral perturbation theory, in \cite{ref:cola_durr} and \cite{ref:beci_villa}.

\subsection*{Acknowledgements}
We thank S.~D\"urr, R.~Sommer and S.~Capitani for many useful discussions.
The computer center at NIC/DESY Zeuthen provided the necessary technical
help and the computer resources. This work was supported by
the EU IHP Network on Hadron Phenomenology from Lattice QCD
and by the DFG Sonderforschungsbereich/Transregio SFB/TR9-03.

\newpage




\def\NPB #1 #2 #3 {Nucl.~Phys.~{\bf#1} (#2)\ #3}
\def\NPBproc #1 #2 #3 {Nucl.~Phys.~B (Proc. Suppl.) {\bf#1} (#2)\ #3}
\def\PRD #1 #2 #3 {Phys.~Rev.~{\bf#1} (#2)\ #3}
\def\PLB #1 #2 #3 {Phys.~Lett.~{\bf#1} (#2)\ #3}
\def\PRL #1 #2 #3 {Phys.~Rev.~Lett.~{\bf#1} (#2)\ #3}
\def\PR  #1 #2 #3 {Phys.~Rep.~{\bf#1} (#2)\ #3}

\def\etal{{\it et al.}}
\def\ibid{{\it ibid}.}

\input refs.tex

\end{document}

%% file: refs.tex
\newpage

\def\NPB #1 #2 #3 {Nucl.~Phys.~{\bf#1} (#2)\ #3}
\def\NPBproc #1 #2 #3 {Nucl.~Phys.~B (Proc. Suppl.) {\bf#1} (#2)\ #3}
\def\PRD #1 #2 #3 {Phys.~Rev.~{\bf#1} (#2)\ #3}
\def\PLB #1 #2 #3 {Phys.~Lett.~{\bf#1} (#2)\ #3}
\def\PRL #1 #2 #3 {Phys.~Rev.~Lett.~{\bf#1} (#2)\ #3}
\def\PR  #1 #2 #3 {Phys.~Rep.~{\bf#1} (#2)\ #3}

\def\etal{{\it et al.}}
\def\ibid{{\it ibid}.}